# A new form of the energy-momentum tensor of the interaction of an electromagnetic field with a non-conducting medium. The wave equations. The electromagnetic forces

Yurii A. Spirichev

Research and Design Institute of Radio-Electronic Engineering - branch of Federal State Unitary Enterprise of Federal Scientific-Production Center "Production Association "Start" named after Michael V.Protsenko"

E-mail: yurii.spirichev@mail.ru

(Dated: April 12, 2017)

**Abstract**

The article describes a new approach to obtaining the energy-momentum tensor of electromagnetic field in medium without the use of Maxwell's equations and Poynting theorem. The energy-momentum tensor has new qualities and consequences. Its linear invariant is Lagrange density of the electromagnetic field. From the tensor follows the equation of conservation of energy density, the equation of flux energy density and wave equations for these energy values. Wave equation for momentum density describes simultaneous transfer of momentum and angular momentum regardless of radiation polarization. From the tensor follow the balance equations of the electromagnetic forces for the momentum density in the forms of the Minkowski and Abraham, which proves their equality and mutual supplementation. Equation for Abraham force is obtained as well. It is shown that its divergence is equal to zero. Tensor and the balance equations of electromagnetic forces in a continuous medium are derived.

**Keywords**: the energy-momentum tensor, Abraham-Minkowski controversy, the electromagnetic forces, Abraham force, wave equations, momentum density, angular momentum.

## The contents

## 1. Introduction

The problem of interaction of electromagnetic fields (EMF) with medium is being discussed for years, but a unique solution has not been found yet. In recent years attempts to create materials with unique electromagnetic values are being undertaken. Therefore, the issue of interaction of EMF with medium has gone to the frontburner. Electromagnetic forces in continuous medium are usually expected to be found in the form of four-dimensional divergence of energy-momentum tensor of (EMT) [1], playing a key role in this task. The problem of finding electromagnetic forces in a continuous medium can be divided into two parts. The first one is to choose the form of EMT. The second one is to choose material equations describing the electromagnetic properties of the medium. The aim of this article is to show the solution to the first problem, which says that there is no definite answer to the question whether one of the known forms of EMT is correct. The most frequently discussed forms of EMT are Minkowski's tensor and Abraham's tensor. For example, this is done, in the articles [1] - [10], [14] – [33]. In the articles [1] - [3], [5], [8], [10] the authors are conducting a comparative analysis of the results following from EMT in the forms of Minkowski and Abraham and give their preference to Abraham's tensor.

In the articles [4], [6], [7], [9] advantages of EMT in the form of Minkowski and disadvantages of EMT in the form of Abraham are shown. According to the authors' opinion in the articles [4] and [6] Abraham's tensor is considered to be non-relativistically covariant, and therefore the preference is given to Minkowski's variant of tensor. In the article [2] it is noted that "in most situations, the results obtained on the basis of the of Abraham's and Minkowski's tensors, are absolutely identical". According to the authors' opinion in the article [3], in the framework of purely macroscopic approach, it is impossible to make an unambiguous choice of EMT form. An important aspect of this debate is the question of the existence of the Abraham force. This force appears when comparing tensors of Minkowski and Abraham as a necessary addition to the Abraham's tensor.

One of the reasons for existence of different points of view on the EMT forms is the lack of strong mathematical method of derivation of the EMT. In the well-known research works it is obtained not by mathematical derivation, but by the method of construction from the different parts. From Maxwell equations and the expression for the Lorentz force with the use of Poynting's theorem they get the equations which are interpreted as equations of conservation of energy and momentum. Further, members of these equations are interpreted as derivatives of the EMT components. These "building" parts of the EMT are the energy density and the momentum density of the EMF, energy flux density (the Poynting vector), a three-dimensional tensor of momentum flux density (or three-dimensional stress tensor). This method has a certain freedom in choosing component parts of EMT and leads to the fact that they, sometimes, are chosen by the authors on the basis of general consideration and understood in different ways. This provokes a debate. This method was used to build EMT in the forms of Minkowski, Hertz – Heaviside, Abraham, Helmholtz - Abraham, Abraham – Brillouin – Pitaevskii, Polevoi – Rytov, etc. These forms of EMT correspond to a certain representation forms of electromagnetic forces. This method of deriving EMT has a significant drawback. The main drawback is that relative dielectric and magnetic permittivity of the medium in the Poynting's theorem are considered to be constant [2], As a result, the obtained results are not universal. In relation to electromagnetic forces in the article [3] it is noted that they are obtained by "consequential method. Meanwhile, it is desirable to obtain force and other quantities (energy density, energy flux, momentum density) with the help of a unified method based on the field equations". The field equations (system of Maxwell's equations) are derived from the antisymmetric tensor of the EMF, so it is methodically reasonable to derive EMT directly from electromagnetic field tensor, omitting the step of obtaining Maxwell's equations, Poynting's theorem and the choice of the component parts of EMT. With this method in hand, the form of the EMT and following from it the equations for energy and momentum are completely and uniquely determined by the antisymmetric tensor of EMF. Thus, no reasons for a debate arise.

This article presents strong consistent mathematical approach for deriving EMT, the equations of conservation of density of electromagnetic energy and momentum density of electromagnetic forces, the wave equations for the density of energy and momentum. The feature of this method is that it does not use Maxwell's equations and the Poynting's theorem but uses fundamental element - antisymmetric tensor of the EMF, which contains all necessary information. The antisymmetric tensor of the EMF itself follows from the four-dimensional derivative of the electromagnetic potential – antisymetric tensor of EMF - through its decomposition into symmetric and antisymmetric tensors. Thus, the only starting point of the proposed method is the four-dimensional electromagnetic potential $\mathbf{A}_\mu$. In the method presented below all the equations of conservation and balance of the electromagnetic forces result from EMT in the form of its four-dimensional divergence. This approach is simple, descriptive, and with proper initial postulate and the absence of mathematical errors gives the correct result. EMT and following equations, derived with the help of this method, are suitable for consideration of the general cases of interaction of EMF with the medium, since no restrictions on the forms of material equations are imposed.

**2. The tensors of the electromagnetic field and electromagnetic induction**

The geometry of space-time is taken as the pseudo-Euclidean Minkowski space (ct, ix, iy, iz) [11]. Four-dimensional electromagnetic potential $\mathbf{A}_\mu$, is respectively, defined as ($\varphi/c$, $i\mathbf{A}$), where $\varphi$ and $\mathbf{A}$ are the scalar and vector potentials of EMF. Asymmetric 4-tensor of the second rank $F_{\nu\mu}$ is derived as four-dimensional derivation of the electromagnetic potential $\mathbf{A}_\mu$:

$$F_{\nu\mu} = \partial_\nu \mathbf{A}_\mu = \partial \mathbf{A}_\mu (\varphi/c, i\mathbf{A}_k) / \partial \mathbf{r}_\nu (c \cdot t, i \cdot \mathbf{r}_k)$$

where **r**$_k$ - coordinates in Euclidean space. The asymmetric tensor of the electromagnetic field Fνμ in matrix representation has the form:

$$F_{\nu\mu} = \partial_\nu \mathbf{A}_\mu = \begin{pmatrix} \frac{1}{c^2}\partial_t\varphi & \frac{1}{c}i\cdot\partial_t A_x & \frac{1}{c}i\cdot\partial_t A_y & \frac{1}{c}i\cdot\partial_t A_z \\ -\frac{1}{c}i\cdot\partial_x\varphi & \partial_x A_x & \partial_x A_y & \partial_x A_z \\ -\frac{1}{c}i\cdot\partial_y\varphi & \partial_y A_x & \partial_y A_y & \partial_y A_z \\ -\frac{1}{c}i\cdot\partial_z\varphi & \partial_z A_x & \partial_z A_y & \partial_z A_z \end{pmatrix} \qquad (1)$$

From this tensor, in the form of its four-dimensional divergence for each of the indices $\partial_\nu F_{\nu\mu} = 0$ and $\partial_\mu F_{\nu\mu} = 0$, follow the equations of EMF in the potentials (it is assumed that the field sources are missing):

$$\frac{1}{c^2}\partial_{tt}\varphi - \Delta\varphi = 0 \qquad \frac{1}{c^2}\partial_{tt}\mathbf{A} - \Delta\mathbf{A} = 0 \qquad \partial_t(\frac{1}{c^2}\partial_t\varphi + \nabla\cdot\mathbf{A}) = 0 \qquad -\nabla(\frac{1}{c^2}\partial_t\varphi + \nabla\cdot\mathbf{A}) = 0$$

The first two equations are the Maxwell's equations in the Lorentz calibration without the sources, and the other two are derivatives of the calibration of the Lorentz condition. By antisymmetrization and symmetrization, tensor (1) can be uniquely decomposed into symmetric and antisymmetric tensors:

$$F_{\nu\mu} = \frac{1}{2}F_{(\nu\mu)} + \frac{1}{2}F_{[\nu\mu]} = \frac{1}{2}(\partial_\nu\mathbf{A}_\mu + \partial_\mu\mathbf{A}_\nu) + \frac{1}{2}(\partial_\nu\mathbf{A}_\mu - \partial_\mu\mathbf{A}_\nu) \qquad (2)$$

In this form antisymmetric tensor of the EMF in the matrix representation has the form:

$$F_{[\nu\mu]} = \begin{pmatrix} 0 & \frac{1}{c}i\cdot(\partial_t A_x + \partial_x\varphi) & \frac{1}{c}i\cdot(\partial_t A_y + \partial_y\varphi) & \frac{1}{c}i\cdot(\partial_t A_z + \partial_z\varphi) \\ -\frac{1}{c}i\cdot(\partial_t A_x + \partial_x\varphi) & 0 & (\partial_x A_y - \partial_y A_x) & (\partial_x A_z - \partial_z A_x) \\ -\frac{1}{c}i\cdot(\partial_t A_y + \partial_y\varphi) & -(\partial_x A_y - \partial_y A_x) & 0 & (\partial_y A_z - \partial_z A_y) \\ -\frac{1}{c}i\cdot(\partial_t A_z + \partial_z\varphi) & -(\partial_x A_z - \partial_z A_x) & -(\partial_y A_z - \partial_z A_y) & 0 \end{pmatrix} = \begin{pmatrix} 0 & -\frac{1}{c}i\cdot E_x & -\frac{1}{c}i\cdot E_y & -\frac{1}{c}i\cdot E_z \\ \frac{1}{c}i\cdot E_x & 0 & B_z & -B_y \\ \frac{1}{c}i\cdot E_y & -B_z & 0 & B_x \\ \frac{1}{c}i\cdot E_z & B_y & -B_x & 0 \end{pmatrix} \qquad (3)$$

where **E** - electric field intensity; **B** - magnetic field induction. From this tensor follow Maxwell's equations for vacuum or microfield.

Zommerfeld A. [12] divided the electromagnetic values into force values and quantity values. Force values: electric field intensity **E** and the magnetic field induction **B**. Quantitative values: the induction of electric field **D** and the magnetic field intensity **H**. Pairs of values **E** and **B**, **D** and **H** can be combine respectively into the antisymmetric tensors of the EMF $F_{[\nu\mu]}$ (3) and electromagnetic induction (EMI) $f_{[\nu\mu]}$ [12]. We write the tensor EMI $f_{[\nu\mu]}$ by analogy to the tensor of EMF (3) in the form:

$$f_{[\nu\mu]} = \begin{pmatrix} 0 & -i\cdot c\cdot D_x & -i\cdot c\cdot D_y & -i\cdot c\cdot D_z \\ i\cdot c\cdot D_x & 0 & H_z & -H_y \\ i\cdot c\cdot D_y & -H_z & 0 & H_x \\ i\cdot c\cdot D_z & H_y & -H_x & 0 \end{pmatrix} \qquad (4)$$

From this tensor Maxwell's equations for continuous media follow.

The relationship between **E** and **D**, **B** and **H** is defined by material equations. For vacuum or microfield $\mathbf{D} = \varepsilon_0 \cdot \mathbf{E}$, and $\mathbf{H} = \mathbf{B}/\mu_0$, where $\varepsilon_0$ and $\mu_0$ are, respectively, electric and magnetic constants. For weak EMF in isotropic non-ferromagnetic dielectric medium without dispersion is usually take the material equation in the form:

$$\mathbf{D} = \varepsilon \cdot \varepsilon_0 \cdot \mathbf{E} \quad \text{и} \quad \mathbf{H} = \mathbf{B}/\mu \cdot \mu_0 \qquad (5)$$

Where $\varepsilon$ and $\mu$ are the relative dielectric permittivity and magnetic permeability of the medium.

### 3. The tensors of energy-momentum in electrodynamics

The canonical EMT in the general form is:

$$T_{\nu\mu} = \begin{bmatrix} W & i\frac{1}{c}\mathbf{S} \\ ic \cdot \mathbf{g} & t_{ik} \end{bmatrix} \qquad (\nu, \mu = 0, 1, 2, 3; \ i, k = 1, 2, 3) \qquad (6)$$

where   **W** – energy density;

**S** – the energy flux density (Umov-Poynting vector);

**g** – the density of momentum;

$t_{ik}$ – density momentum flux tensor (the tension tensor).

The received Minkowski's and Abraham's tensors, which were obtained on the basis of Poynting's theorem have received the greatest prevalence in electrodynamics. Components EMT (6) in the form of Minkowski have the form:

$$W = (\mathbf{E} \cdot \mathbf{D} + \mathbf{H} \cdot \mathbf{B})/2 \qquad \mathbf{S} = \mathbf{E} \times \mathbf{H}$$

$$\mathbf{g}^M = \mathbf{D} \times \mathbf{B} \qquad t_{ik}^M = (E_i D_k + H_i B_k) - \delta_{ik}(\mathbf{E} \cdot \mathbf{D} + \mathbf{H} \cdot \mathbf{B})/2.$$

Components EMT (6) in the form of Abraham have the forms:

$$W = (\mathbf{E} \cdot \mathbf{D} + \mathbf{H} \cdot \mathbf{B})/2 \qquad \mathbf{S} = \mathbf{E} \times \mathbf{H}$$

$$\mathbf{g}^A = \mathbf{E} \times \mathbf{H} \qquad t_{ik}^A = (E_i D_k + E_k D_i + H_i B_k + H_k B_i)/2 - \delta_{ik}(\mathbf{E} \cdot \mathbf{D} + \mathbf{H} \cdot \mathbf{B})/2.$$

After substitution of the components of Abraham's form in EMT (6), it becomes symmetric.

### 4. The derivation of the tensor of energy-momentum tensor of fields and induction

We will derive the EMT (6) directly from the tensors of the EMF (3) and EMI (4) without the involvement of Maxwell's equations and Poynting's theorem. Energy values are the products of the force values and the quantity values [12, p. 11]. The energy of interaction of EMF with the medium is a quadratic form from the tension and induction of electric and magnetic fields. Since the intensity and induction of electric and magnetic fields are the components of the tensors (3) and (4), the quadratic form of their components are components of the EMT. Therefore, we will get EMT in the form of a scalar product of tensors of the EMF (3) and EMI (4).

The scalar product of two tensors of the second rank are computed using the formula [13 p. 308]:

$$P_{\nu\mu} = \sum_{\eta=0}^{\eta=3} a_{\nu\eta} b_{\eta\mu} \qquad \nu, \mu=0, 1, 2, 3$$

By making the following replacement: $a_{\nu\eta} = F_{[\nu\eta]}$ and $b_{\eta\mu} = f_{[\eta\mu]}$ we will obtain:

$$T_{\nu\mu} = F_{[\nu\eta]} f_{[\eta\mu]} \qquad \nu, \eta, \mu=0, 1, 2, 3$$

where the members with the same indices are summed. This formula will let us find the components of EMT (6):

$$T_{00} = E_x D_x + E_y D_y + E_z D_y \qquad T_{01} = i \cdot (E_y H_z - E_z H_y)/c$$
$$T_{11} = E_x D_x - B_z H_z - B_y H_y \qquad T_{02} = i \cdot (E_z H_x - E_x H_z)/c$$
$$T_{22} = E_y D_y - B_z H_z - B_x H_x \qquad T_{03} = i \cdot (E_x H_y - E_y H_x)/c$$
$$T_{33} = E_z D_z - B_y H_y - B_x H_x \qquad T_{10} = ic(B_z D_y - B_y D_z)$$
$$T_{20} = ic(B_x D_z - B_z D_x) \qquad T_{30} = ic(B_y D_x - B_x D_y)$$
$$T_{12} = E_x D_y + B_y H_x \qquad T_{13} = E_x D_z + B_z H_x$$
$$T_{21} = E_y D_x + B_x H_y \qquad T_{23} = E_y D_z + B_z H_y$$
$$T_{31} = E_z D_x + B_x H_z \qquad T_{32} = E_z D_y + B_y H_z$$

These components of EMT can be written as:

$$T_{\nu\mu} = \begin{bmatrix} W & i\frac{1}{c}\mathbf{S} \\ ic \cdot \mathbf{g} & t_{ik} \end{bmatrix} = \begin{bmatrix} \mathbf{E} \cdot \mathbf{D} & i \cdot (\mathbf{E} \times \mathbf{H})/c \\ ic \cdot (\mathbf{D} \times \mathbf{B}) & E_i D_k + B_i H_k - 3\delta_{ik}(\mathbf{B} \cdot \mathbf{H}) \end{bmatrix} \qquad (7)$$

where $W = \mathbf{E} \cdot \mathbf{D}$, $\mathbf{S} = \mathbf{E} \times \mathbf{H}$, $\mathbf{g} = \mathbf{D} \times \mathbf{B}$, $t_{ik} = E_i D_k + B_i H_k - 3\delta_{ik}(\mathbf{B} \cdot \mathbf{H})$ i, k=1, 2, 3

Comparison of the components of EMT with Minkowski's and Abraham's tensors shows that it is close to Minkowski's EMT and differs from it by diagonal components, i.e., the energy density W and the diagonal components of the three-dimensional tensor of flux density of momentum $t_{ik}$. The main topic of discussion in articles [1]-[10] is the kind of momentum density **g** in EMT. In the form of Minkowski momentum density is represented as $g^M = \mathbf{D} \times \mathbf{B}$, but in the form of Abraham it is represented as $g^A = \mathbf{E} \times \mathbf{H}/c^2$. The density of the momentum **g** in EMT (7) has the form identical to Minkowski's. In the general case EMT (7) is asymmetric as well as Minkowski's EMT.

For the medium described by the material equations (5), EMT (7) has a symmetric form:

$$T_{\nu\mu} = \begin{bmatrix} \varepsilon\varepsilon_0 \mathbf{E}^2 & i \cdot (\mathbf{E} \times \mathbf{H})/c \\ i \cdot (\mathbf{E} \times \mathbf{H})/c & \varepsilon\varepsilon_0 \mathbf{E}^2 - 2\mu\mu_0 \mathbf{H}^2 \end{bmatrix} \qquad (8)$$

For vacuum and microfield EMT (7) also has a symmetric form:

$$T_{\nu\mu} = \begin{bmatrix} \varepsilon_0 \mathbf{E}^2 & i \cdot (\mathbf{E} \times \mathbf{B})/c\mu_0 \\ i \cdot (\mathbf{E} \times \mathbf{B})/c\mu_0 & \varepsilon_0 \mathbf{E}^2 - 2\mathbf{B}^2/\mu_0 \end{bmatrix} \qquad (9)$$

Linear invariants EMT of (7) – (9) are

$$I_1 = 2(\mathbf{E}\cdot\mathbf{D} - \mathbf{B}\cdot\mathbf{H}) \quad I_2 = 2(\varepsilon\varepsilon_0\mathbf{E}^2 - \mu\mu_0\mathbf{H}^2) \quad I_3 = 2(\varepsilon_0\mathbf{E}^2 - \mathbf{B}^2/\mu_0)$$

These invariants represent the Lagrange density of the EMF, and a linear invariant $I_3$ for the microfield is also a quadratic invariant of antisymmetric tensor of the EMF (3). Such positive qualities are absent in known forms of EMT.

### 5. The equations for the conservation and wave equations for energy and momentum of the electromagnetic field

The equations of conservation of electromagnetic energy and momentum follow from EMT (7) in the form of its four-dimensional divergence. In general case EMT (7) is asymmetric and for each of its indices two groups of equations can be given (taking into consideration the form of equation of EMT (7), it is possible not to make any distinctions between covariant and contravariant indices):

$$\text{a)} \ \partial_\nu T_{\nu\mu} = 0 \quad \text{and} \quad \text{б)} \ \partial_\mu T_{\nu\mu} = 0$$

or   a) $\frac{1}{c}\partial_t W + c\cdot\nabla\cdot\mathbf{g} = 0 \quad \frac{1}{c^2}\partial_t \mathbf{S} - \partial_i t_{ik} = 0$ and  б) $\partial_t W + \nabla\cdot\mathbf{S} = 0 \quad \partial_t \mathbf{g} - \partial_k t_{ik} = 0$

In the first group we will receive the equations of energy density conservation of EMF and energy flow S:

$$\frac{1}{c}\partial_t(\mathbf{E}\cdot\mathbf{D}) + c\cdot\nabla\cdot(\mathbf{D}\times\mathbf{B}) = 0 \tag{10}$$

$$\frac{1}{c^2}\partial_t(\mathbf{E}\times\mathbf{H}) - \partial_i(E_i D_k + B_i H_k - 3\delta_{ik}(\mathbf{B}\cdot\mathbf{H})) = 0 \tag{11}$$

In the second group we will receive the equations of energy density conservation and momentum density in media g:

$$\frac{1}{c}\partial_t(\mathbf{E}\cdot\mathbf{D}) + \nabla\cdot(\mathbf{E}\times\mathbf{H})/c = 0 \tag{12}$$

$$\partial_t(\mathbf{D}\times\mathbf{B}) - \partial_k(E_i D_k + B_i H_k - 3\delta_{ik}(\mathbf{B}\cdot\mathbf{H})) = 0 \tag{13}$$

From the equations of the energy density conservation (10) and (12) follows the equation:

$$c\cdot\nabla\cdot(\mathbf{D}\times\mathbf{B}) = \nabla\cdot(\mathbf{E}\times\mathbf{H})/c \quad \text{or} \quad \nabla\cdot(\mathbf{D}\times\mathbf{B}) = \nabla\cdot(\mathbf{E}\times\mathbf{H})/c^2 \quad \text{or} \quad \nabla\cdot\mathbf{g}^M = \nabla\cdot\mathbf{g}^A$$

i.e. the divergence momentum density in the forms of the Minkowski and Abraham are equal. Taking derivatives with time from both parts of the last equation, we will obtain:

$$\nabla\cdot\partial_t \mathbf{g}^M = \nabla\cdot\partial_t \mathbf{g}^A \quad \text{or} \quad \nabla\cdot(\partial_t \mathbf{g}^M - \partial_t \mathbf{g}^A) = 0$$

The expression in brackets represents the Abraham force. Consequently, it follows from EMT (7) that the divergence of the Abraham force is equal to zero. This conclusion follows from the Minkowski's tensor. There are no restrictions on constitutive equations in equations (10) – (13). Therefore, equations (10) – (13) are universal and describe the conservation laws of energy density, electromagnetic energy flux density and momentum density in all types of material equations for EMF and EMI. The expression in brackets represents the power of Abraham. Consequently, TEM (7) it follows that the divergence of the Abraham force is equal to zero. This conclusion follows from the Minkowski tensor. In equations (10) – (13), there are no restrictions on constitutive equations. Therefore, equations (10) – (13) are generic and describe the conservation laws of energy density, flux density electromagnetic energy and density of the pulse in all types of material equations for EMF and EMI. The resulting equations were obtained for the stationary medium, but due to relativistic covariance of EMF and EMI tensors, these equations are also covariant, and, when using the known formulas of transition, are valid for a moving medium.

For vacuum or microfield equations (10) and (12) are reduced to one equation

$$\frac{1}{c^2}\partial_t \mathbf{E}^2 + \nabla \cdot (\mathbf{E} \times \mathbf{B}) = 0 \qquad (14)$$

Taking into consideration that the scalar product of the mixed components of the vector is equal to zero, equations (11) and (13) are simplified and also reduced to one equation:

$$\frac{1}{c^2}\partial_t (\mathbf{E} \times \mathbf{B}) - \nabla(\frac{1}{c^2}\mathbf{E}^2 - 2\mathbf{B}^2) = 0 \qquad (15)$$

Expanding the second term of the equation (14) taking into account Maxwell equations, reduce equation to form:

$$\nabla \cdot (\mathbf{E} \times \mathbf{B}) = \mathbf{B} \cdot (\nabla \times \mathbf{E}) - \mathbf{E} \cdot (\nabla \times \mathbf{B}) = -\mathbf{B} \cdot \partial_t \mathbf{B} - \varepsilon_0 \mu_0 \mathbf{E} \cdot \partial_t \mathbf{E} = -\frac{1}{2}\partial_t \mathbf{B}^2 - \frac{1}{2}\frac{1}{c^2}\partial_t \mathbf{E}^2$$

Substituting this expression into equation (14), we will obtain $\partial_t \mathbf{E}^2 / c^2 - \partial_t \mathbf{B}^2 = 0$, or with accuracy to constants we will obtain equation for the electromagnetic wave $\mathbf{E}^2 / c^2 = \mathbf{B}^2$. Taking into account this equation, equation (15) can be written in the form:

$$\partial_t (\mathbf{E} \times \mathbf{B}) + \nabla \mathbf{E}^2 = 0 \qquad (16)$$

Despite the fact that the electromagnetic energy travels in waves, there are no wave equations for EMF energy, energy flow and electromagnetic momentum in electrodynamics. Let us obtain these equations. Considering the equations (14) and (16) as a system, and dividing the unknown quantities in a standard way, we will derive the wave equation for the energy density of the electric field:

$$\frac{1}{c^2}\partial_{tt}\mathbf{E}^2 - \Delta \mathbf{E}^2 = 0 \qquad (17)$$

and the wave equation for flux density S of the electromagnetic energy

$$\frac{1}{c^2}\partial_{tt}(\mathbf{E} \times \mathbf{B}) - \nabla(\nabla \cdot (\mathbf{E} \times \mathbf{B})) = \frac{1}{c^2}\partial_{tt}\mathbf{S} - \nabla(\nabla \cdot \mathbf{S}) = 0 \qquad (18)$$

Taking into account the equation $\mathbf{E}^2 / c^2 = \mathbf{B}^2$ equation (17) gives us the wave equation for the magnetic field energy:

$$\frac{1}{c^2}\partial_{tt}\mathbf{B}^2 - \Delta \mathbf{B}^2 = 0 \qquad (19)$$

Dividing equation (19) on $c^2$, we derive the wave equation for the density of electromagnetic momentum:

$$\frac{1}{c^2}\partial_{tt}\mathbf{g} - \nabla(\nabla \cdot \mathbf{g}) = 0 \qquad (20)$$

Thus, equations (17) - (20) describe the structure of energy and momentum of electromagnetic radiation. Let us consider this question in detail. The wave equation (20) can be written in the form:

$$\frac{1}{c^2}\partial_{tt}\mathbf{g} - \nabla \times \nabla \times \mathbf{g} - \Delta \mathbf{g} = 0 \qquad (21)$$

It is known that electromagnetic radiation simultaneously transfers momentum and angular momentum. Equation (21) describes both of these characteristics of radiation. If we eliminate the term $\nabla \times \nabla \times \mathbf{g}$ from the equation (21), we will obtain the classical D'alembert wave equation, which describes the transfer of momentum density. The second term of the equation (21) describes the double circulation of the momentum density in a closed loop, i.e. it describes the transfer of angular momentum density. In electrodynamics there are different points of view on angular momentum of the EMF, discussed in the articles [34] - [36]. In the article [35] it is noted that in classical electrodynamics the paradox of "null-helicity" of a plane electromagnetic wave exists, when the equations of the EMF do not describe the transfer process of angular momentum of a plane wave, which contradicts the ideas of quantum electrodynamics about the internal angular momentum (spin) of a photon, independent of tradiation polarization. This problem is solved by the equation (21), which shows that the term $\nabla \times \nabla \times \mathbf{g}$, which describes double rotation of the vector of momentum density, describes the transfer of the angular momentum density. This corresponds to quantum electrodynamics understanding of the internal angular momentum (spin) of the photon. Since the vector of momentum density here has dual rotation, it implies that the electromagnetic wave has a toroidal moment [37] of the momentum density. Hence, the internal angular momentum of the photon is also the toroidal angular momentum of the electromagnetic field. Thus, equation (21) eliminates the paradox of "zero helicity" of electromagnetic wave, showing the spirality of motion of electromagnetic energy in it, regardless of radiation polarization and brings the classical electrodynamics and quantum electrodynamics closer to each other.

### 6. Electromagnetic forces in continuous media and their tensor

Electromagnetic forces, more accurately the density of electromagnetic forces in a continuous non-conductive medium, are defined as derivatives of the electromagnetic momentum density at the time $\partial_t \mathbf{g}$. In the absence of external forces, charges and currents, equations (11) and (13) following from EMT (7), can be considered as the balance equations of electromagnetic forces in the media. Equation (11) for the momentum density in the form of Abraham can be written as:

$$\partial_t \mathbf{g}^A = \partial_t (\mathbf{E} \times \mathbf{H})/c^2 = \partial_i (E_i D_k + B_i H_k - 3\delta_{ik} (\mathbf{B} \cdot \mathbf{H})) \qquad (22)$$

Equation (13) for the momentum density in the form of Minkowski can be written in the form:

$$\partial_t \mathbf{g}^M = \partial_t (\mathbf{D} \times \mathbf{B}) = \partial_k (E_i D_k + B_i H_k - 3\delta_{ik} (\mathbf{B} \cdot \mathbf{H})) \qquad (23)$$

From the Minkowski's tensor two similar equations for the momentum density in the forms of Abraham and Minkowski follow, but they are not equal to equations (22) and (23):

$$\partial_t \mathbf{g}^A = \partial_t (\mathbf{E} \times \mathbf{H})/c^2 = \partial_i ((E_i D_k + H_i B_k) - \delta_{ik} (\mathbf{E} \cdot \mathbf{D} + \mathbf{H} \cdot \mathbf{B})/2)$$

$$\partial_t \mathbf{g}^M = \partial_t (\mathbf{D} \times \mathbf{B}) = \partial_k ((E_i D_k + H_i B_k) - \delta_{ik} (\mathbf{E} \cdot \mathbf{D} + \mathbf{H} \cdot \mathbf{B})/2)$$

From the Abraham's tensor follows an equation just for the momentum density in the form of Abraham:

$$\partial_t \mathbf{g}^A = \partial_t (\mathbf{E} \times \mathbf{H})/c^2 = \partial_i ((E_i D_k + E_k D_i + H_i B_k + H_k B_i)/2 - \delta_{ik} (\mathbf{E} \cdot \mathbf{D} + \mathbf{H} \cdot \mathbf{B})/2)$$

The electromagnetic forces in a non-conductive medium are defined by two quantities– induction of the electric field **D** and magnetic field intensity **H,** which respectively depend on the electrical and magnetic characteristics of the media. Then equation (11) with the momentum density in the form of Abraham, which includes the magnetic field intensity **H,** describes the electromagnetic forces associated with magnetic characteristics of the medium, and the equation (13) with the momentum density in the form of Minkowski, which includes the flux density of electric field **D,** describes the electromagnetic forces associated with the electrical characteristics of media. For brevity sake, we will call these the densities of electromagnetic forces respectively, the magnetic and electric forces. Based on this, we can conclude that from EMT (7) and the Minkowski tensor, follow the description of both - electric and magnetic forces in the media, i.e. the electromagnetic forces are described in full, and from EMT in the form of Abraham follows only the description of the magnetic forces. This suggests that the Abraham's tensor is incomplete. In general case electric and magnetic forces have different values, and the difference in these electromagnetic forces is the Abraham force. Since the Abraham's tensor does not include this force, in order to obtain all the forces we have to add this force. In general terms, Abraham force is written as the difference of the equations for the changes of the momentum in Minkowski form and in the form of Abraham [3]:

$$\mathbf{F}_A = \partial_t \mathbf{g}^M - \partial_t \mathbf{g}^A = \partial_t (\mathbf{D} \times \mathbf{B}) - \partial_t (\mathbf{E} \times \mathbf{H})/c^2$$

From equations (22) and (23) it follows that the Abraham force can be written also in the form of a difference of the divergence of the stress tensor $t_{ik}$:

$$\mathbf{F}_A = \partial_t \mathbf{g}^M - \partial_t \mathbf{g}^A = \partial_k t_{ik} - \partial_i t_{ik}$$

or $\quad \mathbf{F}_A = \partial_k (E_i D_k + B_i H_k - 3\delta_{ik} (\mathbf{B} \cdot \mathbf{H})) - \partial_i (E_i D_k + B_i H_k - 3\delta_{ik} (\mathbf{B} \cdot \mathbf{H})) = \nabla \times (\mathbf{E} \times \mathbf{D} + \mathbf{B} \times \mathbf{H})$

Finally, the equation Abraham force can be written as:

$$\mathbf{F}_A = \partial_t (\mathbf{D} \times \mathbf{B}) - \partial_t (\mathbf{E} \times \mathbf{H})/c^2 = \nabla \times (\mathbf{E} \times \mathbf{D} + \mathbf{B} \times \mathbf{H}) \qquad (24)$$

This equation also follows from Minkowski's EMT. Equation (24) confirms the conclusion made in chapter 5 that the divergence of the Abraham force is equal to zero. As in equations (11) and (13), there are no restrictions on constitutive equations, and they are universal for any media, this applies to the equations of electromagnetic forces (22) - (24). From EMT (7) and Minkowski follow an equation of forces for the momentum density in the forms of Minkowski and Abraham, hence the subject of discussion about which of these forms of momentum density is the "right" one, has no meaning, as both forms are correct and complement each other.

From equation (24) follows an important conclusion. If the media is described by the canonical material equations of the form $\mathbf{D} = \varepsilon \cdot \varepsilon_0 \cdot \mathbf{E}$ and $\mathbf{H} = \mathbf{B}/\mu \cdot \mu_0$, $\varepsilon$ and $\mu$ are constants or scalar functions, then the vectors $\mathbf{D}$ and $\mathbf{E}$, $\mathbf{H}$ and $\mathbf{B}$ are collinear and the right side of the equation (24) vector product of these vectors is equal to zero. Then the Abraham force is equal to zero, and EMT (7) and Minkowski's tensor are symmetric. Thus, Abraham force which is commonly written as

$$\mathbf{F}_A = \frac{\varepsilon \cdot \mu - 1}{c^2} \partial_t (\mathbf{E} \times \mathbf{H})$$

equate to zero. For this case, the nondiagonal components of the stress tensor $\mathbf{t}_{ik}$ is equal to zero, and the electromagnetic forces acting on media, are determined only by its diagonal members. Then the equations of electromagnetic forces (22) and (23) can be written in the form of one equation:

$$f = \partial_t \mathbf{g}^M = \partial_t \mathbf{g}^A = \partial_t (\mathbf{E} \times \mathbf{H})/c^2 = \nabla(\varepsilon \cdot \varepsilon_0 \cdot \mathbf{E}^2 - 2\mu \cdot \mu_0 \cdot \mathbf{H}^2) \tag{25}$$

From equation (25) we can conclude that depending on magnitude relation of the values of relative dielectric permittivity and magnetic permeability of the medium the electromagnetic force can change its sign or become zero.

For the case of collinear vectors $\mathbf{D}$ and $\mathbf{E}$, $\mathbf{H}$ and $\mathbf{B}$, when the Abraham force is equal to zero, from Minkowski's and Abraham's tensors follow similar equation for the electromagnetic forces:

$$f = \partial_t \mathbf{g}^M = \partial_t \mathbf{g}^A = \partial_t (\mathbf{E} \times \mathbf{H})/c^2 = \nabla(\varepsilon \cdot \varepsilon_0 \cdot \mathbf{E}^2 + \mu \cdot \mu_0 \cdot \mathbf{H}^2/2) \tag{26}$$

Comparison of this equation with the equation (25) shows their principal difference.

Four-dimensional electromagnetic forces are defined as four-dimensional derivatives of EMT. The force balance equations (22) and (23) are obtained in the form of a divergence of EMT. But the electromagnetic force can be obtained in a more general way as the components of a tensor of electromagnetic forces (EFT). To obtain this tensor we will take the four-dimensional derivative of EMT (7). Since differentiation increases tensor rank, the EFT is a third-rank tensor:

$$F_{\eta\nu\mu} = \partial_\eta \mathbf{T}_{\nu\mu} = \partial_\eta \begin{bmatrix} W & i\frac{1}{c}\mathbf{S} \\ ic \cdot \mathbf{g} & \mathbf{t}_{ik} \end{bmatrix} = \partial_\eta \begin{bmatrix} \mathbf{E} \cdot \mathbf{D} & i \cdot (\mathbf{E} \times \mathbf{H})/c \\ ic \cdot (\mathbf{D} \times \mathbf{B}) & E_i D_k + B_i H_k - 3\delta_{ik}(\mathbf{B} \cdot \mathbf{H}) \end{bmatrix} \tag{27}$$

For a better understanding of electromagnetic forces the components of EFT can be written in the form of their four-dimensional derivatives:

$$\partial_\eta W = \partial_\eta (\mathbf{E} \cdot \mathbf{D}) = \frac{1}{c}\partial_t(\mathbf{E} \cdot \mathbf{D}) - i\nabla(\mathbf{E} \cdot \mathbf{D})$$

$$\partial_\eta (i\frac{1}{c}\mathbf{S}) = \partial_\eta (i \cdot (\mathbf{E} \times \mathbf{H})/c) = i \cdot \frac{1}{c^2}\partial_t(\mathbf{E} \times \mathbf{H}) + \frac{1}{c}\nabla(\mathbf{E} \times \mathbf{H})$$

$$\partial_\eta (ic \cdot \mathbf{g}) = \partial_\eta (ic \cdot (\mathbf{D} \times \mathbf{B})) = i \cdot \partial_t(\mathbf{D} \times \mathbf{B}) + c \cdot \nabla(\mathbf{D} \times \mathbf{B})$$

$$\partial_\eta t_{ik} = \partial_\eta (E_i D_k + B_i H_k - 3\delta_{ik}(\mathbf{B} \cdot \mathbf{H})) = \frac{1}{c}\partial_t(E_i D_k + B_i H_k - 3\delta_{ik}(\mathbf{B} \cdot \mathbf{H})) - i\nabla(E_i D_k + B_i H_k - 3\delta_{ik}(\mathbf{B} \cdot \mathbf{H}))$$

Obtained equations for electromagnetic forces are universal for any media, described by the tensor of electromagnetic induction.

Let us derive equations for electromagnetic forces in the media described by the material equations (5):

$$\partial_\eta W = \partial_\eta (\varepsilon\varepsilon_0 \mathbf{E}^2) = \frac{1}{c}\partial_t(\varepsilon\varepsilon_0 \mathbf{E}^2) - i\nabla(\varepsilon\varepsilon_0 \mathbf{E}^2)$$

$$\partial_\eta (ic \cdot \mathbf{g}) = \partial_\eta (i\frac{1}{c}\mathbf{S}) = \partial_\eta (i \cdot (\mathbf{E} \times \mathbf{H})/c) = \frac{1}{c^2}\partial_t(i \cdot (\mathbf{E} \times \mathbf{H})) + \nabla(\mathbf{E} \times \mathbf{H})/c$$

$$\partial_\eta t_{ik} = \partial_\eta (\varepsilon\varepsilon_0 \mathbf{E}^2 - 2\mu\mu_0 \mathbf{H}^2) = \frac{1}{c}\partial_t(\varepsilon\varepsilon_0 \mathbf{E}^2) - 2\frac{1}{c}\partial_t(\mu\mu_0 \mathbf{H}^2) - i \cdot \nabla(\varepsilon\varepsilon_0 \mathbf{E}^2 - 2\mu\mu_0 \mathbf{H}^2)$$

Let us find four-dimensional divergence in each of the pairs of indices of EFT (27) and we will obtain balance equations of electromagnetic forces in a continuous medium in the absence of external forces:

$$\partial_\eta \partial_\nu F_{\eta\nu\mu} = 0 \qquad \partial_\eta \partial_\mu F_{\eta\nu\mu} = 0 \qquad \partial_\nu \partial_\mu F_{\eta\nu\mu} = 0$$

All equations, obtained from EMT (7), are true for a stationary medium, but because of its relativistic covariance, when using the known formulas of transition, they are also valid for a uniformly moving medium.

**7. Conclusion**

From the tensors of the EMF and EMI without the involvement of Maxwell's equations and Poynting's theorem we've strictly mathematically derived EMT (7) from which follow the equations of conservation of electromagnetic energy density, energy flux density and momentum density.

EMT (7) has an important feature that is not observed in other forms of EMT, namely, that its linear invariant presents itself as a quadratic invariant of EMF tensor and Lagrangian density of EMF at the same time, combining these fundamental energy values of the EMF.

New wave equations for electromagnetic energy density, energy flux density and momentum density, that didn't exist in electrodynamics before, were obtained from EMT (7).

The wave equation for momentum density describes energy structure of electromagnetic radiation and simultaneous transfer of the electromagnetic momentum and angular momentum, regardless of the polarization of the radiation. This eliminates the problem of "null-helicity" of plane electromagnetic wave and draws classical electrodynamics and quantum electrodynamics closer to each other.

Obtained balance equations of electromagnetic forces in a continuous medium lead to the conclusion about the equality and mutual complementation of Minkowski's and Abraham's forms of momentum density. This allows to end the discussion on this issue. The equation Abraham force (previously absent in the electrodynamics) is obtained from EMT (7). It is shown that Abraham force exists only in the medium where the vectors **D** and **E**, **H** and **B** are not collinear. Tensor of electromagnetic forces is obtained from EMT (7).

Possibility to derive equations for electromagnetic energy density, energy flux density and momentum density from EMT (7) proves its correctness. Obtaining these equations from the known forms of EMTs is impossible.